\documentclass[aps,prd,showpacs,twocolumn,
superscriptaddress]{revtex4-1}

\usepackage{graphicx}
\usepackage{subfigure} 
\usepackage{dcolumn}

\graphicspath{{figures/}}
\usepackage{amsmath}
\usepackage{amssymb}
\usepackage{bm}
\usepackage{color}
\usepackage{comment}

\usepackage[normalem]{ulem}
\usepackage[dvipsnames]{xcolor}
\usepackage{hyperref}
\usepackage{multirow}

\hypersetup{
colorlinks=true, 
linkcolor=blue,  
citecolor=cyan,  
}

\newcommand{\bea}{\begin{eqnarray}}
\newcommand{\eea}{\end{eqnarray}}
\newcommand{\be}{\begin{equation}}
\newcommand{\ee}{\end{equation}}
\graphicspath{{figures/}}

\begin{document}

\title{Inferring the cosmological constant in early Universe only by gravitational waves}

\author{Song Li}
\email{leesong@shao.ac.cn}
\affiliation{Shanghai Astronomical Observatory, Shanghai, 200030, China }
\affiliation{School of Astronomy and Space Science, University of Chinese Academy of Sciences,
Beijing, 100049, China}

\author{Wen-Biao Han}
\email{wbhan@shao.ac.cn}
\affiliation{Shanghai Astronomical Observatory, Shanghai, 200030, China }
\affiliation{School of Fundamental Physics and Mathematical Sciences, Hangzhou Institute for Advanced Study, UCAS, Hangzhou 310024, China }
\affiliation{School of Astronomy and Space Science, University of Chinese Academy of Sciences,
Beijing, 100049, China}
\affiliation{International Centre for Theoretical Physics Asia-Pacific, Beijing/Hangzhou, China}
\affiliation{Shanghai Frontiers Science Center for  Gravitational Wave Detection, 800 Dongchuan Road, Shanghai 200240, China}

\date{\today}

\begin{abstract}
The expansion of the Universe is accelerating which can be interpreted as due to the cosmological constant $\Lambda$. In this study, we investigate the behavior of gravitational waves in the presence of a cosmological constant at the early universe. We rigorously analyze the merger rate of binary primordial black holes (PBHs) and the corresponding signal-to-noise ratio within the framework of Laser Interferometer Space Antenna (LISA). We find that binary PBHs with a total mass of $M_{\mathrm{tot}}=1000M_{\odot}$ and a redshift larger than $z=500$ are the ideal system for studying the effect of the cosmological constant through LISA. By computing the fisher information matrix, we establish that the cosmological constant can be effectively constrained.
\end{abstract}

\maketitle

\section{Introduction}
In the past decade, significant advancements have been made in the study of gravitational waves (GWs). A major breakthrough occurred in 2015 when the LIGO-Virgo Collaboration detected GW150914, marking the first observation of gravitational wave events from a binary black hole system\cite{GW150914}. Subsequently, the LIGO-Virgo-KAGRA Collaboration has reported the discovery of more compact binary systems. Over 90 of these systems have been identified to date, including two neutron star and black hole systems that are distinct from other binary black hole systems\cite{GW_events_1, GW_events_2, GW_events_3, GW_events_4}. These unique compact binaries present excellent opportunities to test general relativity (GR)\cite{Test_GR_1, Test_GR_2, Test_GR_3, Test_GR_4, Test_GR_5, Test_GR_6} and gain new insights into compact objects\cite{CO_1}, potentially leading to the discovery of new theories beyond GR. Ground-based detectors such as LIGO, Virgo, and KAGRA have been crucial in advancing our understanding of compact binary physics and astrophysics. Looking ahead, future space-based detectors like the Laser Interferometer Space Antenna (LISA)\cite{LISA_des}, Taiji\cite{taiji}, and Tianqin\cite{tianqin} will provide new opportunities to enhance our comprehension of the Universe further.

The recent NANOGrav experiments on pulsar timing arrays reported, for the first time in 2023, evidence for a signal from the stochastic gravitational wave background (SGWB) at nHz frequencies\cite{NANOGrav_1, NANOGrav_2}. In the next decade, the space-based gravitational wave detectors LISA and Taiji will be launched and operative to observe signals at the mHz frequencies. Besides receiving many signals from different astrophysical sources, LISA and Taiji will be listening for possible signals in the SGWB generated at the electroweak scale\cite{SGWB_LISA_1, SGWB_LISA_2}.

The discovery that the expansion of the Universe is accelerating, and there are many possible mechanisms have been proposed to account for this phenomenon, such as the modified gravity\cite{MG_1, MG_2, MG_3}, higher dimensions\cite{HD} and so on. Among them, the cosmological constant is generally believed to be the simplest explanation\cite{CC_1, CC_2}. Cosmological constant is expected to be relevant across all physical scales. For example, the cosmological constant may influence gravitational lensing and contribute to the gravitational equilibrium of large astrophysical structures. Thus, a fundamental question arises: how does the cosmological constant affect gravitational waves? 

It is well known that gravitational waves can serve as a ``standard siren'' for studying cosmology alongside their electromagnetic counterparts\cite{Cosmology_1, Cosmology_2, Cosmology_3, Cosmology_4}. However, when lacking electromagnetic counterparts, the degeneration of luminosity distance and redshift in the flat Friedmann-Lema\^itre-Robertson-Walker (FLRW) spacetime renders the study of cosmology with gravitational waves infeasible. Nevertheless, this degeneration appears in the flat spacetime background, and in principle should be broken by considering the curved background. 

In this study, we analyze gravitational waves in spacetime background characterized by a non-zero cosmological constant, $\Lambda$, within the realm of perturbation theory involving de Sitter (dS) and anti-de Sitter (AdS) metrics. Previous studies have generally concluded that the influence of the cosmological constant on gravitational waves is minimal. However, our research focuses on PBHs in the early universe which will enhance the effect of the cosmological constant on gravitational waves. The paper is organized in the following way. In Sec.~\ref{Merger_Rate_sec} the merger rate of PBH in different configurations is discussed. Section \ref{Waveform_sec} is devoted to describing the gravitational waveform with the cosmological constant and briefly discusses the relation between the Hubble parameter and the cosmological constant. Then we calculate the signal-to-noise ratio of the primordial black hole (PBH) binary with different parameters and the Fisher information matrix is discussed. Finally, Sec.~\ref{Conclusion_sec} summarizes the main results obtained and the conclusions that can be drawn given future observations.

\section{Merger Rate}\label{Merger_Rate_sec}
This section will calculate the event rate of binary mergers involving Primordial Black Holes (PBHs). The detailed information can be found in \cite{Misao_16}. For simplicity's sake, we assume that all PBHs have the same mass.

Firstly, we define $f$ as the fraction of Primordial Black Holes in the dark matter, that is, $\Omega_{\rm BH}=f \Omega_{\rm DM}$. The physical mean separation of BHs, denoted as ${\bar x}$, during matter-radiation equality at redshift $z=z_{\rm eq}$ can then be determined by:
\be
{\bar x}={\left( \frac{M_{\rm BH}}{\rho_{\rm BH}(z_{\rm eq})} \right)}^{1/3}
=\frac{1}{(1+z_{\rm eq}) f^{1/3}} 
{\left( \frac{8\pi G}{3H_0^2} \frac{M_{\rm BH}}{\Omega_{\rm DM}} \right)}^{1/3}.
\ee
Let us consider two neighboring black holes (BHs) that are separated by a physical distance $x$ during matter-radiation equality. The condition for this pair to potentially form a binary system is given by $x$. In this study, we impose this condition on the distance $x$.

In a realistic scenario, there are other black holes present, and the third nearest BH to the BH pair exerts a tidal force that influences the infall motion of the BHs in the pair. Consequently, a head-on collision does not occur, and the BHs in the pair typically form a binary system with a high eccentricity. The major and minor axes of the binary at the time of formation are denoted by $a$ and $b$, respectively.
\be
a=\frac{\alpha}{f} \frac{x^4}{{\bar x}^3},~~~~~b=\beta {\left( \frac{x}{y} \right)}^3 a, 
\label{ellipse}
\ee
where $y$ represents the physical distance to the third black hole at $z=z_{\rm eq}$, while $\alpha$ and $\beta$ denote numerical factors of order ${\cal O}(1)$. From a comprehensive investigation of binary formation dynamics, it is suggested that $\alpha = 0.4$ and $\beta = 0.8$. However, for simplicity in the subsequent analysis, we adopt $\alpha = \beta = 1$.

The eccentricity of the binary at the formation time is given by
\be
e=\sqrt{1-{\left( \frac{x}{y} \right)}^6}.
\label{eccentricity}
\ee

We assume a uniform probability distribution for both $x$ and $y$ in three-dimensional space. Therefore, the probability $dP$ that the potential binary black holes have a separation in the range $(x,x+dx)$ and the distance to the perturbing BH falls within the range $(y,y+dy)$ is given by:
\be
dP=\frac{9}{{\bar x}^6} x^2 y^2 dx dy.
\label{P_i}
\ee 

After performing some calculations, Eq.~\ref{P_i} can be transformed from the $x-y$ plane to the $t-e$ plane. By integrating over $e$ for a fixed value of $t$, the probability of coalescence occurring within the time interval $(t, t+dt)$ can be obtained in the following form:
\be
dP_t=
    \frac{3}{58} \bigg[ -{\left( \frac{t}{T} \right)}^{3/8}
+{\left( \frac{t}{T} \right)}^{3/37} \bigg] \frac{dt}{t}
  \label{dpt}
\ee
where $T=\frac{\bar{x}Q}{f^4}$, $Q=\frac{3}{170}(\frac{GM_{\mathrm{BH}}}{c^{5/3}})^{-3}$. The probability of coalescence occurring within the time interval $(0, t)$ can be expressed simply as $P_c(t)=\int_0^t dP_t$.

The current average number density of Primordial Black Holes (PBHs), denoted as $n_{\rm BH}$, can be calculated using the following formula:
\be
n_{\rm BH}=\frac{3H_0^2}{8\pi G} \frac{\Omega_{\rm BH}}{M_{\rm BH}}.
\ee
Then, the event rate becomes
\be
\begin{aligned} 
{\rm R}
&=n_{\rm BH}\lim_{\Delta t \to 0} \frac{P_c(t_0)-P_c(t_0-{\Delta t})}{\Delta t}
&=\frac{3H_0^2}{8\pi G} \frac{\Omega_{\rm BH}}{M_{\rm BH}} \frac{dP_c}{dt}\bigg|_{t_0}. \label{event_rate}
\end{aligned} 
\ee
where $t_0$ is the age of the Universe.

\begin{figure*}
\centering
\includegraphics[width=0.45 \textwidth]{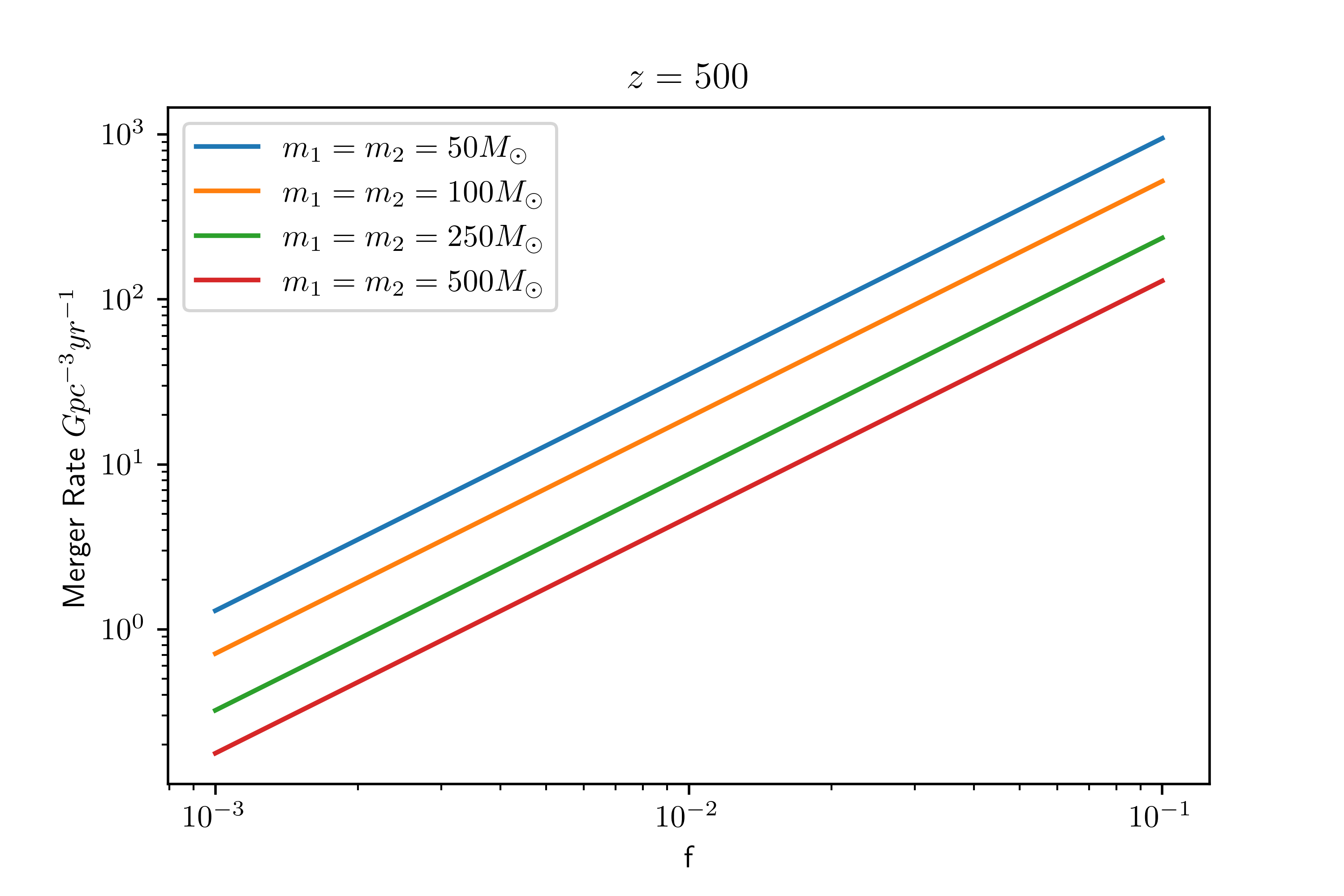}
\includegraphics[width=0.45 \textwidth]{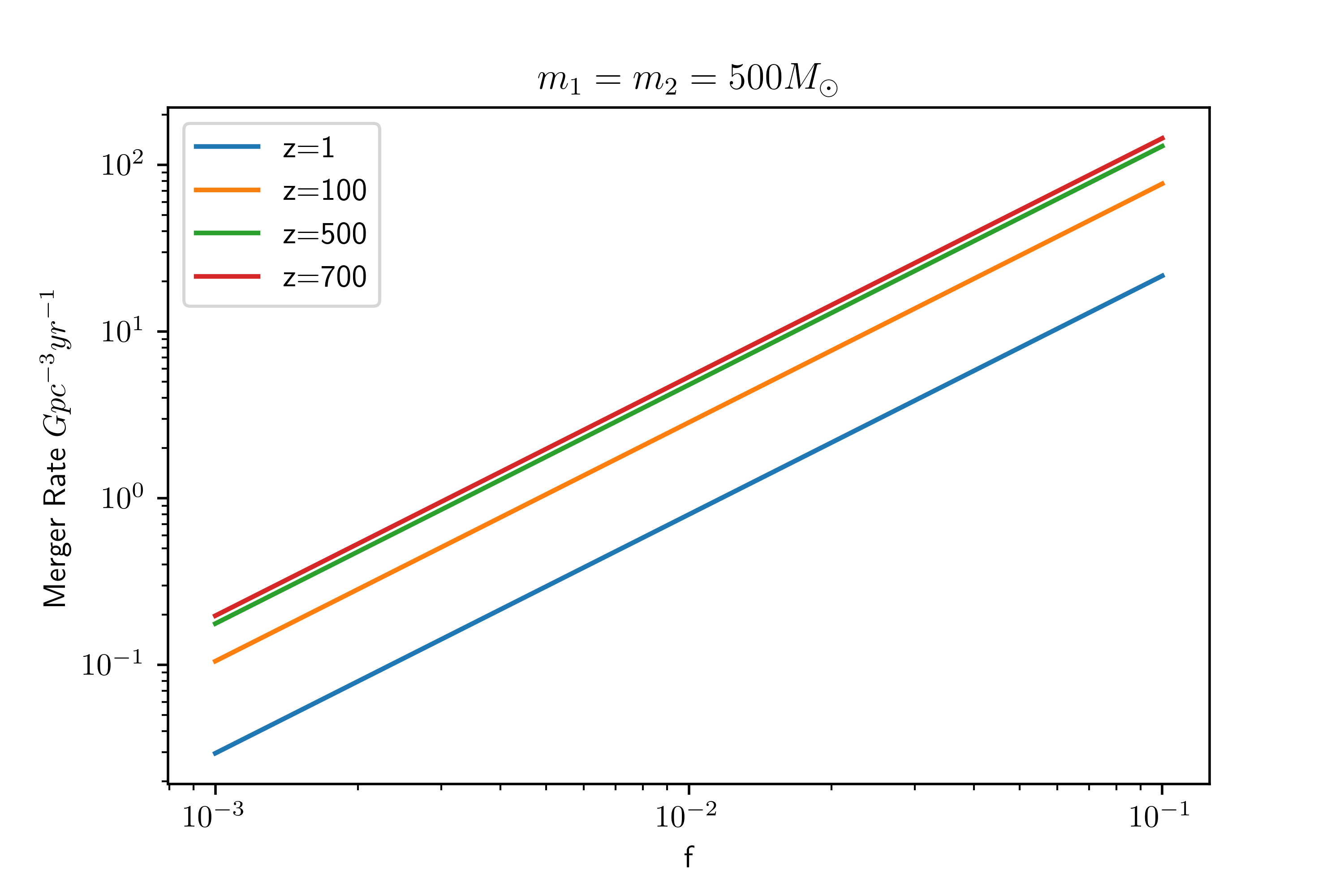}
\caption{Merger rate for different redshift and mass of PBH binaries as a function of the PBH fraction in dark matter $f=\Omega_{\mathrm{BH}}/\Omega_{\mathrm{DM}}$. The left panel shows the merger rate of different masses of PBH binaries at redshift $z=500$, and the right panel shows the merger rate of $500M_{\odot}$-$500M_{\odot}$ PBH binaries for different redshift.}\label{Merger_Rate}
\end{figure*}

Fig.~\ref{Merger_Rate} depicts the merger rate of Primordial Black Hole (PBH) binaries at different redshifts and masses as a function of the PBH fraction $\mathrm{f}=\Omega_{\mathrm{BH}}/\Omega_{\mathrm{DM}}$ in the dark matter. Here, we adopt the latest values of $\Omega_{\mathrm{DM}}=0.27$ and $H_0=70\mathrm{km\,Mpc^{-1}\,s^{-1}}$. In the left panel, it is observed that a fixed redshift $z$ leads to a decrease in the merger rate with increasing PBH mass. Furthermore, an increase in $\mathrm{f}$ corresponds to an increase in the merger rate, as a greater PBH fraction in dark matter results in more PBHs. The right panel demonstrates that higher redshift values correspond to higher merger rates, as the earlier universe possesses a greater abundance of PBHs. However, as $z$ reaches a sufficiently high value, the change in the merger rate becomes negligible.

\section{Gravitational Waveform With Cosmological Constant}\label{Waveform_sec}
In this section, we present the gravitational waveform with the cosmological constant. The most straightforward solution for the waveform with the cosmological constant, derived from the vacuum Einstein equations, has already been established in previous research\cite{Lambda_09}. Building upon their findings, we can now compute the gravitational waveform with the cosmological constant for a circular orbit:
\begin{equation}
\begin{aligned}
\tilde{h}_{+}(f)=&A\left(\frac{\Lambda R_{f}^{3} f}{36 c}+1\right)^{-\frac{1}{16}} e^{i \Psi_{+}(f)} \frac{c}{r}\left(\frac{G M_{c}}{c^{3}}\right)^{5 / 6} \\
&\frac{1}{f^{7 / 6}}\left(\frac{1+\cos ^{2} \iota}{2}\right)
\label{Lambda_GW_amplitude}
\end{aligned}
\end{equation}

with the phase,
\begin{equation}
\begin{aligned}
\Psi_{+}(f)=&2 \pi f\left(t_{c}+r / c\right)-\Phi_{c}-\frac{\pi}{4}+\frac{3}{4}\left(\frac{G M_{c}}{c^{3}} 8 \pi f\right)^{-5 / 3}\\ 
&\left(\frac{\Lambda R_{f}^{3} f}{3 c}+1\right)^{-1}
\label{Lambda_GW_phase}
\end{aligned}
\end{equation}

where $\Lambda$ represents the cosmological constant, $R_f$ denotes the observation time, $M_c$ is defined as the chirp mass using the formula $\frac{(m_1m_2)^{3/5}}{(m_1+m_2)^{1/5}}$, $c$ represents the speed of light, $G$ symbolizes the gravitational constant, $r$ and $\iota$ represent the distance and angle between the observer and source, and $t_c$ and $\phi_c$ represent the collision time and collision phase. We observe that Eqs.~(\ref{Lambda_GW_amplitude})-(\ref{Lambda_GW_phase}) contain more physical information compared to Equation (31) in \cite{Lambda_09}. Hence, we extend these two equations to a phenomenological model\cite{PhD} in order to obtain a more general set of equations for our analysis.

The phase ansatz in the inspiral stage of the phenomenological model is expressed as follows:
\begin{equation}
\begin{aligned}
\phi_{\mathrm{Ins}}= & \phi_{\mathrm{TF} 2}(M f ; \Xi) \\
& +\frac{1}{\eta}\left(\sigma_{0}+\sigma_{1} f+\frac{3}{4} \sigma_{2} f^{4 / 3}+\frac{3}{5} \sigma_{3} f^{5 / 3}+\frac{1}{2} \sigma_{4} f^{2}\right)
\label{PhD_inspiral}
\end{aligned}
\end{equation}

where $\eta=m_1 m_2/M^2$, $M = m_1+m_2$, the $\phi_{\mathrm{TF} 2}$ is the full TaylorF2 phase:
\begin{equation}
\begin{aligned}
\phi_{\mathrm{TF} 2}= & 2 \pi f t_{c}-\varphi_{c}-\pi / 4 \\
& +\frac{3}{128 \eta}(\pi f M)^{-5 / 3} \sum_{i=0}^{7} \varphi_{i}(\Xi)(\pi f M)^{i / 3}
\end{aligned}
\end{equation}
The constants $\sigma_{i}$ (where $i = 0, 1, 2, 3, 4$) represent the correlation between the mass and spin of the system, while $\varphi_{i}(\Xi)$ denotes the PN expansion coefficients associated with the intrinsic binary parameters. The detailed information of $\sigma_{i}$ and $\varphi_{i}(\Xi)$ can be found in Appendix B of this article \cite{PhD}.

We will utilize Eqs.~(\ref{Lambda_GW_amplitude})-(\ref{PhD_inspiral}) to derive the gravitational waveform incorporating the cosmological constant. However, it is crucial to determine the values of $\Lambda$ that signify the presence of cosmological constants. A study by \cite{Lambda_09} revealed that a gravitational wave with $\Lambda=10^{-52}$ cannot be distinguished by LIGO. Hence, in this study, we explore the possibility of increasing the value of $\Lambda$ to achieve a more significant discernible difference. According to the Friedmann equations, there exists a relationship between the Hubble parameter $H$ and the scale factor $a$:

\begin{equation}
\begin{aligned}
H\left ( t \right ) \equiv \frac{\dot{a}\left ( t \right )   }{{a}\left ( t \right )   } 
\end{aligned}
\end{equation}

the scale factor $a$ has a different relationship with time:
\begin{equation}
\begin{aligned}
a\left ( t \right ) \propto \left\{\begin{matrix}
  t^{1/2}&\textrm{Radiation-dominated era} \\
  t^{2/3}&\textrm{Matter-dominated era} \\
  \mathrm{exp}\left ( H_0t \right ) &\textrm{Dark-energy-dominated era}
\end{matrix}\right.
\end{aligned}
\end{equation}

where $H_0$ is the Hubble constant. 

From the above equations and $\Lambda\propto H^2/c^2$, we can get the relationship between cosmological constant $\Lambda$ and time $t$:
\begin{equation}
\Lambda\propto \frac{1}{ct} \label{Lambda_t_equ}
\end{equation}

\begin{figure}
\centering
\includegraphics[width=0.40 \textwidth]{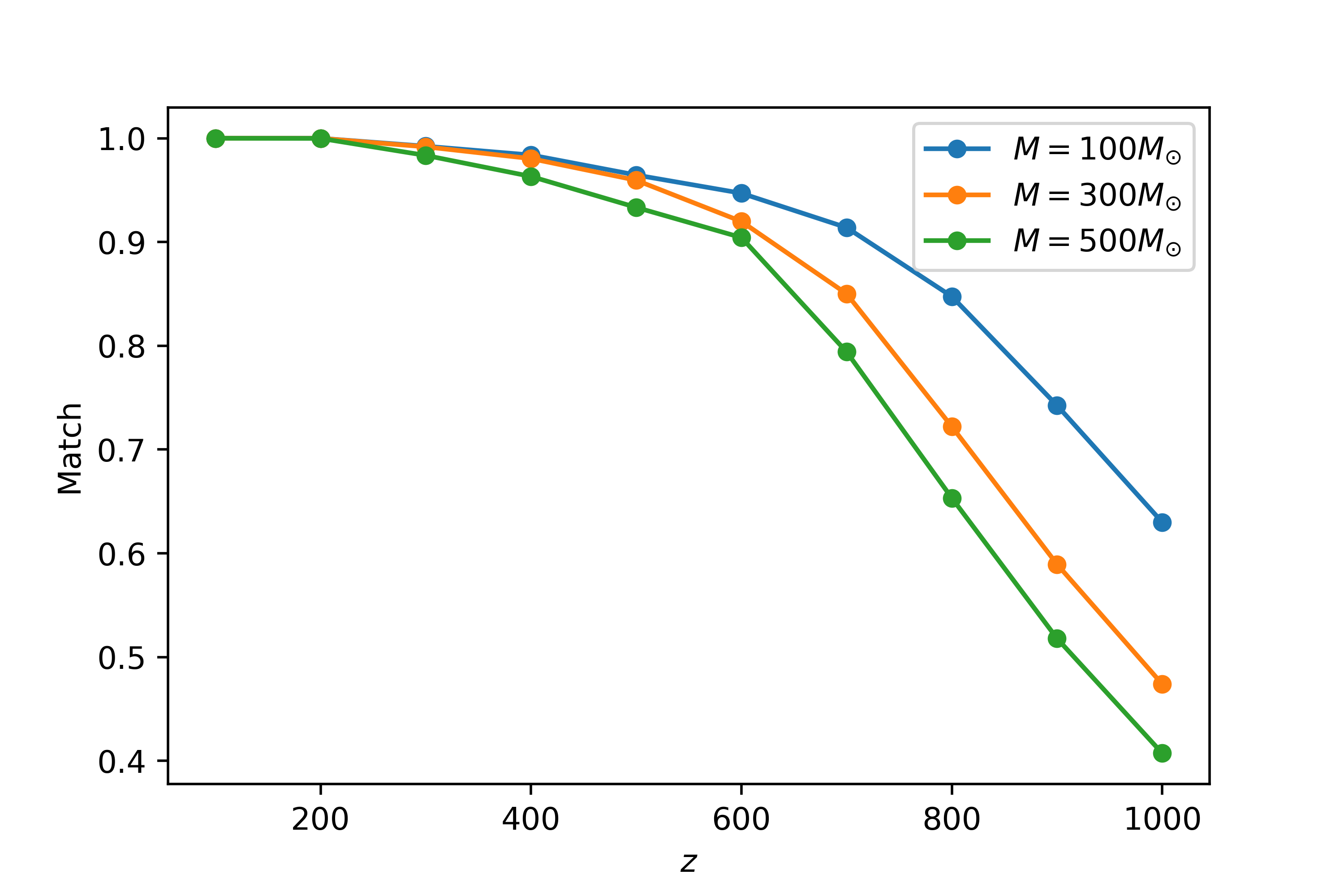}
\caption{The match between the waveforms with cosmological constant $\Lambda$ and without it at different redshift. The different colors represent the different masses of the PBHs.}\label{Match_fig}
\end{figure}

According to Eq.~\ref{Lambda_t_equ}, the value of the cosmological constant increases as time decreases. Therefore, the cosmological constant $\Lambda$ exhibits a higher value at a greater redshift, facilitating the distinction of its impact on gravitational waves. Furthermore, we employ the parameter overlap to ensure a more intuitive comparison. The definition of the overlap is as follows:

\begin{equation}
\mathrm{F}= \left[\frac{\left\langle h_{1} \mid h_{2}\right\rangle}{\sqrt{\left\langle h_{1} \mid h_{1}\right\rangle\left\langle h_{2} \mid h_{2}\right\rangle}}\right],
\end{equation}

\begin{equation}
\left\langle h_{1}, h_{2}\right\rangle=4 \operatorname{Re} \int_{f_{\min }}^{f_{\max }} \frac{\tilde{h}_{1}(f) \tilde{h}_{2}^{*}(f)}{S_{n}(f)} d f,
\end{equation}

where $h_1$ represents the waveform incorporating the cosmological constant, $h_2$ denotes the comparative waveform without the cosmological constant, and $S_n(f)$ stands for the power spectral density of the detector noise. In this study, we adopt the sensitivity curve of LISA\cite{LISA_des}. The match is defined as

\begin{equation}
\mathrm{FF}=\max \left[\frac{\left\langle h_{1} \mid h_{2}\right\rangle}{\sqrt{\left\langle h_{1} \mid h_{1}\right\rangle\left\langle h_{2} \mid h_{2}\right\rangle}}\right],
\end{equation}

and the mismatch of two waveforms is defined as $1-\mathrm{FF}$. Fig.~\ref{Match_fig} illustrates the comparison of waveforms with and without a cosmological constant $\Lambda$ at various redshifts. It is evident that as the mass increases, the level of similarity decreases. Hence, it is advisable to focus on primordial black holes with higher masses to observe more significant distinctions. However, as depicted in the left panel of Fig.~\ref{Merger_Rate}, a higher mass correlates with a decrease in the merger rate. Consequently, we opt for a PBH mass of $500M_{\odot}$ to strike a balance between the merger rate and the match. Additionally, it is observed that for redshifts exceeding 500, the similarity drops below 0.95 at $M=500M_{\odot}$, leading us to investigate PBHs at a redshift of $z=500$.

Then we will calculate the signal-to-noise ratio (SNR) for various configurations of binary primordial black holes, including different redshifts, masses, and other parameters.
    
The SNR, $\rho$ can be expressed as:
\begin{equation}{\label{eq:snrint2}}
\rho^{2}=\int_{0}^{\infty} \frac{(2|\tilde{h}(f)| \sqrt{f})^{2}}{S_{n}(f)} \mathrm{d} \ln (f)
\end{equation}
where $\tilde{h}(f)$ is the Fourier transform of the signal, $S_{n}(f)$ is the noise power spectral density (PSD) of the LISA\cite{LISA_des}. The complete derivation is provided in section 7.3 of \cite{Maggiore2007}.

There is already a toolkit available\cite{gwent} for calculating the signal-to-noise ratio of various detector designs, including pulsar timing arrays, space-based detectors, and ground-based detectors, for different binary black hole (BBH) sources. In this study, we utilize this toolkit to evaluate the detectability of different BBH source configurations by the Laser Interferometer Space Antenna (LISA).

\begin{figure*}
\centering
\includegraphics[width=0.45 \textwidth]{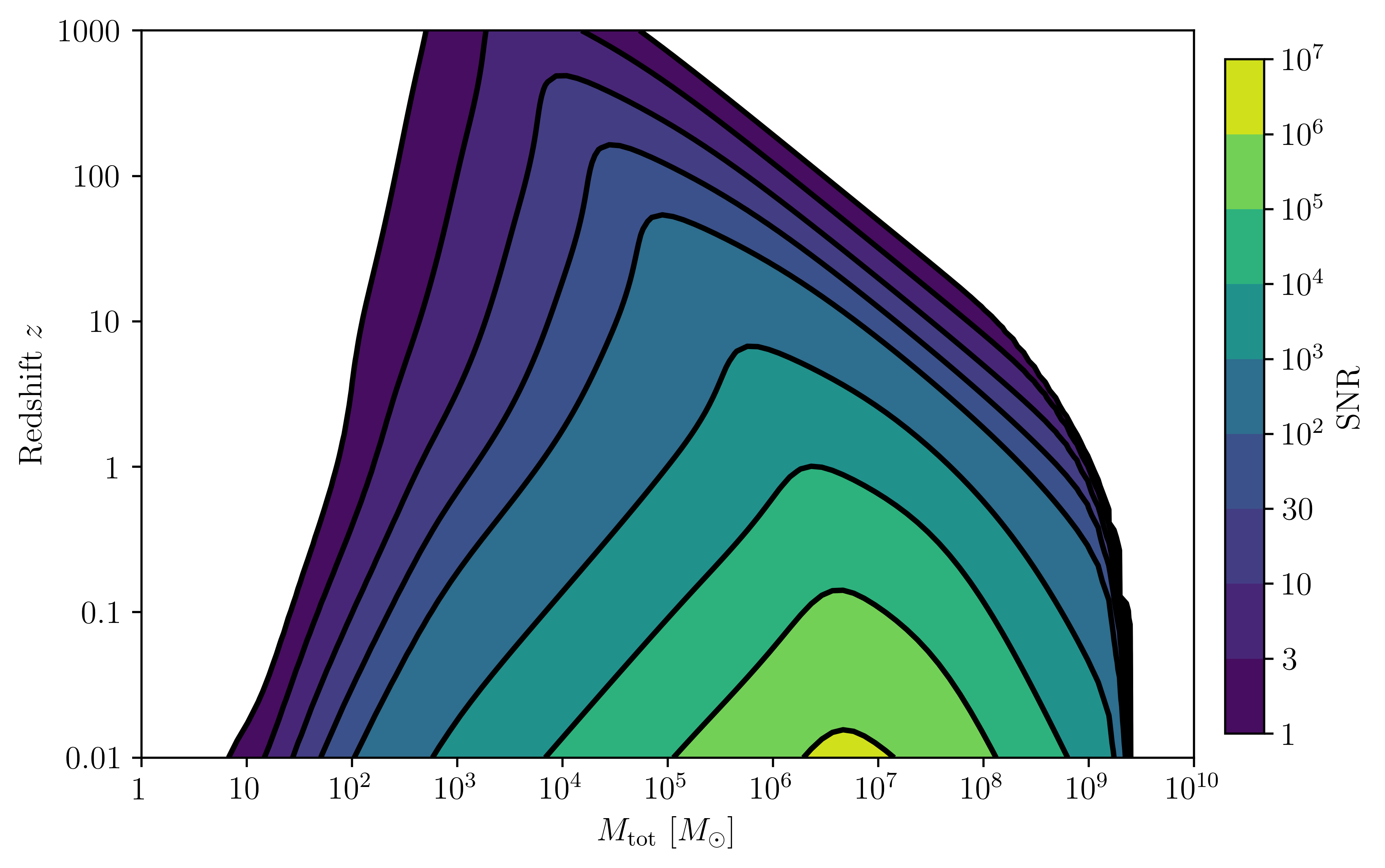}
\includegraphics[width=0.45 \textwidth]{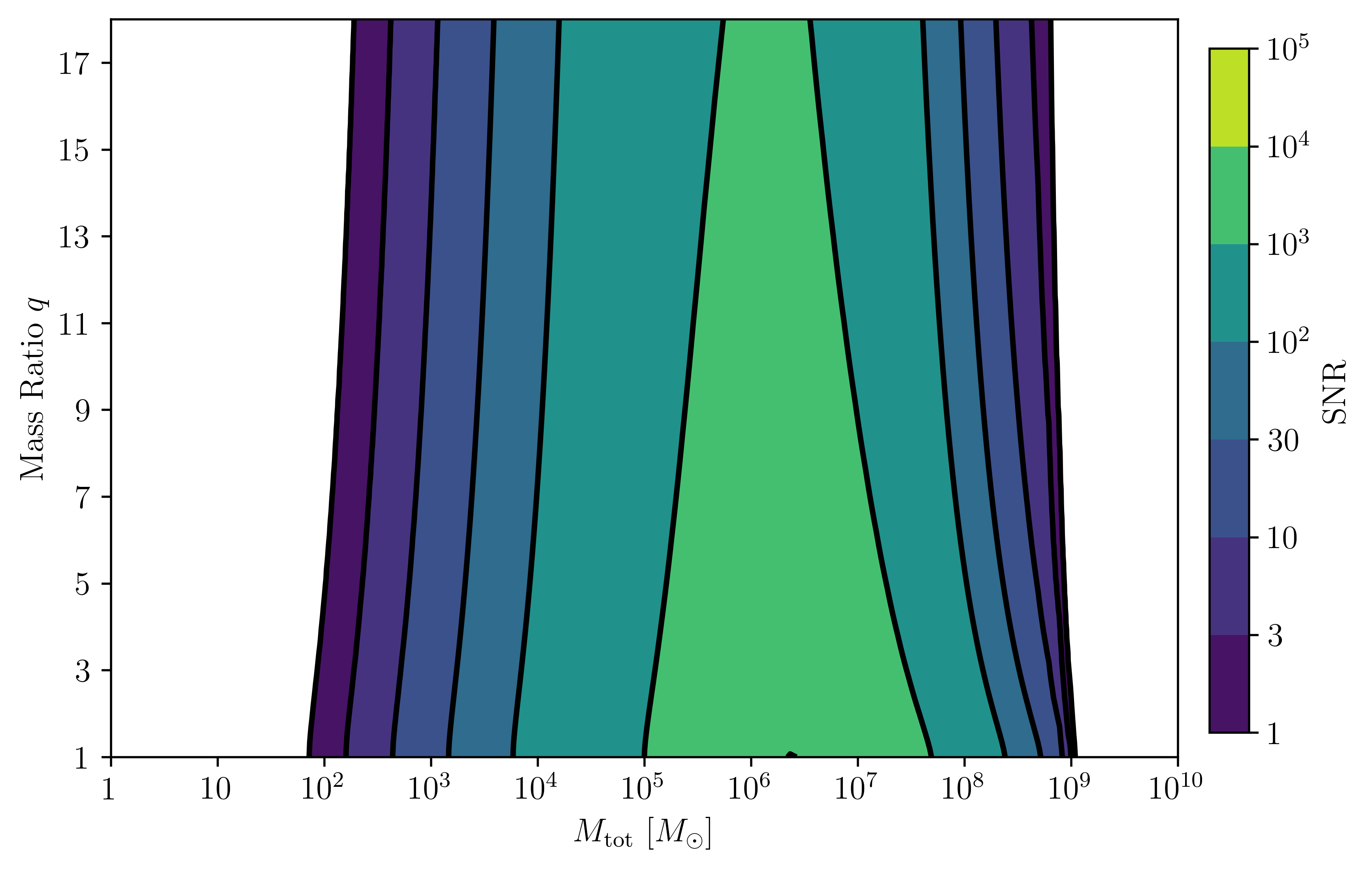}
\caption{The SNR for different mass, redshift, and mass ratio. The left panel shows the SNR for different masses $M_{\mathrm{tot}}$ and redshift $z$. The right panel shows the SNR for different masses $M_{\mathrm{tot}}$ and mass ratios $q$.}\label{SNR}
\end{figure*}

Fig.~\ref{SNR} displays the signal-to-noise ratio for various configurations of PBHs. In the left panel, the SNR is depicted for different combinations of total mass $M_{\mathrm{tot}}$ and redshift $z$. It is evident that in order to achieve a high SNR (SNR $>$ 5) at high redshifts, both $M_{\mathrm{tot}}$ and $z$ need to be within the range of $10^3M_{\odot}$-$10^5M_{\odot}$ and $z > 500$, respectively. The right panel exhibits the SNR for different mass ratios $q$ when the total mass $M_{\mathrm{tot}}$ is fixed. It can be observed that a higher mass ratio corresponds to a lower SNR, albeit the difference is small. Consequently, for the purpose of this research, we opt for two black holes with equal masses. Based on the aforementioned findings, we have selected the configurations of primordial black holes to be studied as $M_{\mathrm{tot}}=1000M_{\odot}$, $z=500$, and $q=1$ in the subsequent analysis.

We also calculate the parameter uncertainties. The parameter uncertainties can be approximated by taking the square root of the diagonal elements of the inverse of $\Gamma_{ij}$, i.e., $\Delta \lambda_i \approx \sqrt{(\Gamma^{-1})_{ii}}$. To perform the parameter estimation, we utilize the Fisher information matrix (FIM) method. The FIM for a frequency domain gravitational wave (GW) signal $\tilde{h}(f)$ parameterized by $\lambda$ can be found in \cite{FIM}.

\begin{eqnarray}
\Gamma_{i j} & = & \left\langle\left.\frac{\partial \tilde{h}}{\partial \lambda_{i}} \right\rvert\, \frac{\partial \tilde{h}}{\partial \lambda_{j}}\right\rangle
\end{eqnarray}
with

\begin{eqnarray}
\left\langle\tilde{h}_{1} \mid \tilde{h}_{2}\right\rangle & = & 2 \operatorname{Re} \int_{0}^{\infty} \frac{\tilde{h}_{1}^{*}(f) \tilde{h}_{2}+\tilde{h}_{1}(f) \tilde{h}_{2}^{*}(f)}{S_{n}(f)} \mathrm{d} f
\end{eqnarray}

where $\tilde{h}(f)$ is the frequency-domain waveform, ${S_{n}(f)}$ is the noise power spectral density of different detectors, $\lambda$ is the waveform parameter. Now, we consider the PBH binary with $m_1=m_2=500M_{\odot}$, $a_1=a_2=0.5$, and $\Lambda=10^{-45}$ to get the FIM. The results are presented in Table.~\ref{Fisher}. The first and second row shows the FIM for LISA and Taiji, and the third row shows the combined observations of Taiji and LISA. We note that all the parameters have good estimates. The FIM has the best result when considering the combined observations of Taiji and LISA. We also calculate the probability distribution:

\begin{eqnarray}
p\left(\Delta \lambda^{i}\right)=\mathcal{N} e^{-\frac{1}{2} \Gamma_{i j} \Delta \lambda^{i} \Delta \lambda^{j}},
\end{eqnarray}

where $\mathcal{N}=\sqrt{\mathrm{det}\left ( \Gamma /2\pi \right ) }$ is the appropriate normalization factor.

\begin{table*}[]
\renewcommand{\arraystretch}{1.3}
\resizebox{\textwidth}{!}{
\begin{tabular}{|c|c|c|c|c|c|c|c|l}
\cline{1-8}
     & $\Delta m_1/m_1$                                   & $\Delta m_2/m_2$                                   & $\Delta a_1/a_1$                                   & $\Delta a_2/a_2$                                   & $\Delta t_c/t_c$                                   & $\Delta \phi_c/\phi_c$                             & $\Delta \Lambda/\Lambda$                           &  \\ \cline{1-8}
LISA & $1.43\times 10^{-3}$ & $1.42\times 10^{-3}$ & $2.14\times 10^{-3}$ & $2.14\times 10^{-3}$ & $1.34\times 10^{-3}$ & $1.18\times 10^{-2}$ & $7.11\times 10^{-3}$ &  \\ \cline{1-8}
Taiji & $1.04\times 10^{-3}$ & $1.03\times 10^{-3}$ & $1.56\times 10^{-3}$ & $1.56\times 10^{-3}$ & $1.04\times 10^{-3}$ & $9.27\times 10^{-3}$ & $5.67\times 10^{-3}$ &  \\ \cline{1-8}
LISA-Taiji & $8.40\times 10^{-4}$ & $8.35\times 10^{-4}$ & $1.26\times 10^{-3}$ & $1.26\times 10^{-3}$ & $8.22\times 10^{-4}$ & $7.29\times 10^{-3}$ & $4.43\times 10^{-3}$ &  \\ \cline{1-8}
\end{tabular}
}
\caption{The Fisher information matrix of the PBH binary for different detectors. The first and second row shows the FIM for LISA and Taiji, the third one is a combined observation of Taiji and LISA. The different columns show the results of mass $m_{1(2)}$, spin $a_{1(2)}$, collision time $t_c$, collision phase $\phi_c$, and cosmological constant $\Lambda$.}
\label{Fisher}
\end{table*}


\section{Conclusion}\label{Conclusion_sec}
The cosmological constant provides the simplest explanation for the expansion of the Universe. It influences not just the expansion of the universe, but also phenomena such as gravitational lensing and gravitational waves. In cosmological studies, the value of the cosmological constant $\Lambda$ is determined by the relationship of the cosmological constant to the Hubble Constant $H_0$ and the dark energy density $\Omega_{\Lambda}$. The influence of the cosmological constant on gravitational waves allows for the development of a gravitational waveform model that can be utilized to study and determine the value of the cosmological constant.

In this study, we investigate the gravitational waveform in the presence of a cosmological constant, denoted as $\Lambda$. Previous research has indicated that it is challenging to distinguish the effects of the cosmological constant when its value is $\Lambda=10^{-52}$. However, our work explores the possibility that in the early universe, the value of $\Lambda$ could be higher than $10^{-52}$, thereby enabling us to study the impact of the cosmological constant through gravitational waves. Since our focus is on studying gravitational waves at high redshifts, we concentrate on primordial black holes.  Fig.~\ref{Merger_Rate} displays the merger rate for PBH binaries, revealing a higher quantity of PBH binaries at high redshifts and low masses. We then calculate the match between the waveforms with the cosmological constant $\Lambda$ and without it at various redshifts, as depicted in  Fig.~\ref{Match_fig}. Interestingly, we observe that as the mass increases, the match decreases. Taking into consideration both the merger rate and the match, we select the mass of PBH binaries to be $500M_{\odot}$ and a redshift $z=500$ for our analysis.

We focused on the space-based detector LISA to investigate high redshift gravitational wave events.  Fig.~\ref{SNR} displays the signal-to-noise ratio for different primordial black hole (PBH) binaries, and we observed that the configuration with a mass of $500M_{\odot}$ and a redshift of $z=500$ yields a high SNR (SNR $>$ 5). This finding affirms the feasibility of studying this particular configuration in our research. In Section~\ref{Waveform_sec}, we computed the Fisher matrix for the PBH binary with $m_1=m_2=500M_{\odot}$, $a_1=a_2=0.5$, and $\Lambda=10^{-45}$. As summarized in Table.~\ref{Fisher}, the obtained restrictions on the various parameters are favorable.

In the future, as space-based gravitational wave detectors like LISA and Taiji come into operation, they will enable the detection of gravitational events occurring at high redshifts. We will have the opportunity to investigate the cosmological constant through a different approach than previous studies. We have demonstrated that by combining observations from Taiji and LISA, we can effectively constrain the value of the cosmological constant. This result not only indicates the potential to study the cosmological constant through gravitational waves but also to explore the structure of the universe with it.


\begin{acknowledgments}
This work is supported by The National Key R\&D Program of China (Grant No. 2021YFC2203002), NSFC (National Natural Science Foundation of China) No. 12473075, No. 12173071 and No. 11773059. This work made use of the High Performance Computing Resource in the Core Facility for Advanced Research Computing at Shanghai Astronomical Observatory.
\end{acknowledgments}

\bibliographystyle{apsrev4-1}  

\begin{thebibliography}{35}%
\makeatletter
\providecommand \@ifxundefined [1]{%
 \@ifx{#1\undefined}
}%
\providecommand \@ifnum [1]{%
 \ifnum #1\expandafter \@firstoftwo
 \else \expandafter \@secondoftwo
 \fi
}%
\providecommand \@ifx [1]{%
 \ifx #1\expandafter \@firstoftwo
 \else \expandafter \@secondoftwo
 \fi
}%
\providecommand \natexlab [1]{#1}%
\providecommand \enquote  [1]{``#1''}%
\providecommand \bibnamefont  [1]{#1}%
\providecommand \bibfnamefont [1]{#1}%
\providecommand \citenamefont [1]{#1}%
\providecommand \href@noop [0]{\@secondoftwo}%
\providecommand \href [0]{\begingroup \@sanitize@url \@href}%
\providecommand \@href[1]{\@@startlink{#1}\@@href}%
\providecommand \@@href[1]{\endgroup#1\@@endlink}%
\providecommand \@sanitize@url [0]{\catcode `\\12\catcode `\$12\catcode
  `\&12\catcode `\#12\catcode `\^12\catcode `\_12\catcode `\%12\relax}%
\providecommand \@@startlink[1]{}%
\providecommand \@@endlink[0]{}%
\providecommand \url  [0]{\begingroup\@sanitize@url \@url }%
\providecommand \@url [1]{\endgroup\@href {#1}{\urlprefix }}%
\providecommand \urlprefix  [0]{URL }%
\providecommand \Eprint [0]{\href }%
\providecommand \doibase [0]{http://dx.doi.org/}%
\providecommand \selectlanguage [0]{\@gobble}%
\providecommand \bibinfo  [0]{\@secondoftwo}%
\providecommand \bibfield  [0]{\@secondoftwo}%
\providecommand \translation [1]{[#1]}%
\providecommand \BibitemOpen [0]{}%
\providecommand \bibitemStop [0]{}%
\providecommand \bibitemNoStop [0]{.\EOS\space}%
\providecommand \EOS [0]{\spacefactor3000\relax}%
\providecommand \BibitemShut  [1]{\csname bibitem#1\endcsname}%
\let\auto@bib@innerbib\@empty
\bibitem [{\citenamefont {Abbott}\ \emph
  {et~al.}(2016{\natexlab{a}})\citenamefont {Abbott} \emph
  {et~al.}}]{GW150914}%
  \BibitemOpen
  \bibfield  {author} {\bibinfo {author} {\bibfnamefont {B.~P.}\ \bibnamefont
  {Abbott}} \emph {et~al.} (\bibinfo {collaboration} {LIGO Scientific
  Collaboration and Virgo Collaboration}),\ }\href {\doibase
  10.1103/PhysRevLett.116.061102} {\bibfield  {journal} {\bibinfo  {journal}
  {Phys. Rev. Lett.}\ }\textbf {\bibinfo {volume} {116}},\ \bibinfo {pages}
  {061102} (\bibinfo {year} {2016}{\natexlab{a}})}\BibitemShut {NoStop}%
\bibitem [{\citenamefont {Abbott}\ \emph
  {et~al.}(2016{\natexlab{b}})\citenamefont {Abbott} \emph
  {et~al.}}]{GW_events_1}%
  \BibitemOpen
  \bibfield  {author} {\bibinfo {author} {\bibfnamefont {B.~P.}\ \bibnamefont
  {Abbott}} \emph {et~al.} (\bibinfo {collaboration} {LIGO Scientific
  Collaboration and Virgo Collaboration}),\ }\href {\doibase
  10.1103/PhysRevLett.116.241103} {\bibfield  {journal} {\bibinfo  {journal}
  {Phys. Rev. Lett.}\ }\textbf {\bibinfo {volume} {116}},\ \bibinfo {pages}
  {241103} (\bibinfo {year} {2016}{\natexlab{b}})}\BibitemShut {NoStop}%
\bibitem [{\citenamefont {Abbott}\ \emph
  {et~al.}(2019{\natexlab{a}})\citenamefont {Abbott} \emph
  {et~al.}}]{GW_events_2}%
  \BibitemOpen
  \bibfield  {author} {\bibinfo {author} {\bibfnamefont {B.~P.}\ \bibnamefont
  {Abbott}} \emph {et~al.} (\bibinfo {collaboration} {LIGO Scientific
  Collaboration and Virgo Collaboration}),\ }\href {\doibase
  10.1103/PhysRevX.9.031040} {\bibfield  {journal} {\bibinfo  {journal} {Phys.
  Rev. X}\ }\textbf {\bibinfo {volume} {9}},\ \bibinfo {pages} {031040}
  (\bibinfo {year} {2019}{\natexlab{a}})}\BibitemShut {NoStop}%
\bibitem [{\citenamefont {Venumadhav}\ \emph {et~al.}(2020)\citenamefont
  {Venumadhav}, \citenamefont {Zackay}, \citenamefont {Roulet}, \citenamefont
  {Dai},\ and\ \citenamefont {Zaldarriaga}}]{GW_events_3}%
  \BibitemOpen
  \bibfield  {author} {\bibinfo {author} {\bibfnamefont {T.}~\bibnamefont
  {Venumadhav}}, \bibinfo {author} {\bibfnamefont {B.}~\bibnamefont {Zackay}},
  \bibinfo {author} {\bibfnamefont {J.}~\bibnamefont {Roulet}}, \bibinfo
  {author} {\bibfnamefont {L.}~\bibnamefont {Dai}}, \ and\ \bibinfo {author}
  {\bibfnamefont {M.}~\bibnamefont {Zaldarriaga}},\ }\href {\doibase
  10.1103/PhysRevD.101.083030} {\bibfield  {journal} {\bibinfo  {journal}
  {Phys. Rev. D}\ }\textbf {\bibinfo {volume} {101}},\ \bibinfo {pages}
  {083030} (\bibinfo {year} {2020})}\BibitemShut {NoStop}%
\bibitem [{\citenamefont {Abbott}\ \emph
  {et~al.}(2021{\natexlab{a}})\citenamefont {Abbott} \emph
  {et~al.}}]{GW_events_4}%
  \BibitemOpen
  \bibfield  {author} {\bibinfo {author} {\bibfnamefont {R.}~\bibnamefont
  {Abbott}} \emph {et~al.} (\bibinfo {collaboration} {LIGO Scientific
  Collaboration and Virgo Collaboration}),\ }\href {\doibase
  10.1103/PhysRevX.11.021053} {\bibfield  {journal} {\bibinfo  {journal} {Phys.
  Rev. X}\ }\textbf {\bibinfo {volume} {11}},\ \bibinfo {pages} {021053}
  (\bibinfo {year} {2021}{\natexlab{a}})}\BibitemShut {NoStop}%
\bibitem [{\citenamefont {Okounkova}(2020)}]{Test_GR_1}%
  \BibitemOpen
  \bibfield  {author} {\bibinfo {author} {\bibfnamefont {M.}~\bibnamefont
  {Okounkova}},\ }\href {\doibase 10.1103/PhysRevD.102.084046} {\bibfield
  {journal} {\bibinfo  {journal} {Phys. Rev. D}\ }\textbf {\bibinfo {volume}
  {102}},\ \bibinfo {pages} {084046} (\bibinfo {year} {2020})}\BibitemShut
  {NoStop}%
\bibitem [{\citenamefont {Isi}\ \emph {et~al.}(2019)\citenamefont {Isi},
  \citenamefont {Giesler}, \citenamefont {Farr}, \citenamefont {Scheel},\ and\
  \citenamefont {Teukolsky}}]{Test_GR_2}%
  \BibitemOpen
  \bibfield  {author} {\bibinfo {author} {\bibfnamefont {M.}~\bibnamefont
  {Isi}}, \bibinfo {author} {\bibfnamefont {M.}~\bibnamefont {Giesler}},
  \bibinfo {author} {\bibfnamefont {W.~M.}\ \bibnamefont {Farr}}, \bibinfo
  {author} {\bibfnamefont {M.~A.}\ \bibnamefont {Scheel}}, \ and\ \bibinfo
  {author} {\bibfnamefont {S.~A.}\ \bibnamefont {Teukolsky}},\ }\href {\doibase
  10.1103/PhysRevLett.123.111102} {\bibfield  {journal} {\bibinfo  {journal}
  {Phys. Rev. Lett.}\ }\textbf {\bibinfo {volume} {123}},\ \bibinfo {pages}
  {111102} (\bibinfo {year} {2019})}\BibitemShut {NoStop}%
\bibitem [{\citenamefont {Abbott}\ \emph
  {et~al.}(2019{\natexlab{b}})\citenamefont {Abbott} \emph
  {et~al.}}]{Test_GR_3}%
  \BibitemOpen
  \bibfield  {author} {\bibinfo {author} {\bibfnamefont {B.~P.}\ \bibnamefont
  {Abbott}} \emph {et~al.} (\bibinfo {collaboration} {LIGO Scientific
  Collaboration and Virgo Collaboration}),\ }\href {\doibase
  10.1103/PhysRevLett.123.011102} {\bibfield  {journal} {\bibinfo  {journal}
  {Phys. Rev. Lett.}\ }\textbf {\bibinfo {volume} {123}},\ \bibinfo {pages}
  {011102} (\bibinfo {year} {2019}{\natexlab{b}})}\BibitemShut {NoStop}%
\bibitem [{\citenamefont {Nair}\ \emph {et~al.}(2019)\citenamefont {Nair},
  \citenamefont {Perkins}, \citenamefont {Silva},\ and\ \citenamefont
  {Yunes}}]{Test_GR_4}%
  \BibitemOpen
  \bibfield  {author} {\bibinfo {author} {\bibfnamefont {R.}~\bibnamefont
  {Nair}}, \bibinfo {author} {\bibfnamefont {S.}~\bibnamefont {Perkins}},
  \bibinfo {author} {\bibfnamefont {H.~O.}\ \bibnamefont {Silva}}, \ and\
  \bibinfo {author} {\bibfnamefont {N.}~\bibnamefont {Yunes}},\ }\href
  {\doibase 10.1103/PhysRevLett.123.191101} {\bibfield  {journal} {\bibinfo
  {journal} {Phys. Rev. Lett.}\ }\textbf {\bibinfo {volume} {123}},\ \bibinfo
  {pages} {191101} (\bibinfo {year} {2019})}\BibitemShut {NoStop}%
\bibitem [{\citenamefont {Abbott}\ \emph
  {et~al.}(2019{\natexlab{c}})\citenamefont {Abbott} \emph
  {et~al.}}]{Test_GR_5}%
  \BibitemOpen
  \bibfield  {author} {\bibinfo {author} {\bibfnamefont {B.~P.}\ \bibnamefont
  {Abbott}} \emph {et~al.} (\bibinfo {collaboration} {The LIGO Scientific
  Collaboration and the Virgo Collaboration}),\ }\href {\doibase
  10.1103/PhysRevD.100.104036} {\bibfield  {journal} {\bibinfo  {journal}
  {Phys. Rev. D}\ }\textbf {\bibinfo {volume} {100}},\ \bibinfo {pages}
  {104036} (\bibinfo {year} {2019}{\natexlab{c}})}\BibitemShut {NoStop}%
\bibitem [{\citenamefont {Abbott}\ \emph
  {et~al.}(2021{\natexlab{b}})\citenamefont {Abbott} \emph
  {et~al.}}]{Test_GR_6}%
  \BibitemOpen
  \bibfield  {author} {\bibinfo {author} {\bibfnamefont {R.}~\bibnamefont
  {Abbott}} \emph {et~al.} (\bibinfo {collaboration} {LIGO Scientific
  Collaboration and Virgo Collaboration}),\ }\href {\doibase
  10.1103/PhysRevD.103.122002} {\bibfield  {journal} {\bibinfo  {journal}
  {Phys. Rev. D}\ }\textbf {\bibinfo {volume} {103}},\ \bibinfo {pages}
  {122002} (\bibinfo {year} {2021}{\natexlab{b}})}\BibitemShut {NoStop}%
\bibitem [{\citenamefont {Abbott}\ \emph
  {et~al.}(2019{\natexlab{d}})\citenamefont {Abbott} \emph {et~al.}}]{CO_1}%
  \BibitemOpen
  \bibfield  {author} {\bibinfo {author} {\bibfnamefont {B.~P.}\ \bibnamefont
  {Abbott}} \emph {et~al.},\ }\href {\doibase 10.3847/2041-8213/ab3800}
  {\bibfield  {journal} {\bibinfo  {journal} {The Astrophysical Journal
  Letters}\ }\textbf {\bibinfo {volume} {882}},\ \bibinfo {pages} {L24}
  (\bibinfo {year} {2019}{\natexlab{d}})}\BibitemShut {NoStop}%
\bibitem [{\citenamefont {Amaro-Seoane}\ \emph {et~al.}(2017)\citenamefont
  {Amaro-Seoane} \emph {et~al.}}]{LISA_des}%
  \BibitemOpen
  \bibfield  {author} {\bibinfo {author} {\bibnamefont {Amaro-Seoane}} \emph
  {et~al.},\ }\href {\doibase 10.48550/arXiv.1702.00786} {\bibfield  {journal}
  {\bibinfo  {journal} {arXiv e-prints}\ ,\ \bibinfo {eid} {arXiv:1702.00786}}
  (\bibinfo {year} {2017})},\ \Eprint {http://arxiv.org/abs/1702.00786}
  {arXiv:1702.00786 [astro-ph.IM]} \BibitemShut {NoStop}%
\bibitem [{\citenamefont {{Huang}}\ \emph {et~al.}(2017)\citenamefont
  {{Huang}}, \citenamefont {{Gong}}, \citenamefont {{Xu}}, \citenamefont
  {{Amaro-Seoane}}, \citenamefont {{Bian}}, \citenamefont {{Chen}},
  \citenamefont {{Chen}}, \citenamefont {{Fang}}, \citenamefont {{Feng}},
  \citenamefont {{Liu}}, \citenamefont {{Li}}, \citenamefont {{Li}},
  \citenamefont {{Luo}}, \citenamefont {{Shao}}, \citenamefont {{Spurzem}},
  \citenamefont {{Tang}}, \citenamefont {{Wang}}, \citenamefont {{Wang}},
  \citenamefont {{Zang}},\ and\ \citenamefont {{Lau}}}]{taiji}%
  \BibitemOpen
  \bibfield  {author} {\bibinfo {author} {\bibfnamefont {S.}~\bibnamefont
  {{Huang}}}, \bibinfo {author} {\bibfnamefont {X.}~\bibnamefont {{Gong}}},
  \bibinfo {author} {\bibfnamefont {P.}~\bibnamefont {{Xu}}}, \bibinfo {author}
  {\bibfnamefont {P.}~\bibnamefont {{Amaro-Seoane}}}, \bibinfo {author}
  {\bibfnamefont {X.}~\bibnamefont {{Bian}}}, \bibinfo {author} {\bibfnamefont
  {Y.}~\bibnamefont {{Chen}}}, \bibinfo {author} {\bibfnamefont
  {X.}~\bibnamefont {{Chen}}}, \bibinfo {author} {\bibfnamefont
  {Z.}~\bibnamefont {{Fang}}}, \bibinfo {author} {\bibfnamefont
  {X.}~\bibnamefont {{Feng}}}, \bibinfo {author} {\bibfnamefont
  {F.}~\bibnamefont {{Liu}}}, \bibinfo {author} {\bibfnamefont
  {S.}~\bibnamefont {{Li}}}, \bibinfo {author} {\bibfnamefont {X.}~\bibnamefont
  {{Li}}}, \bibinfo {author} {\bibfnamefont {Z.}~\bibnamefont {{Luo}}},
  \bibinfo {author} {\bibfnamefont {M.}~\bibnamefont {{Shao}}}, \bibinfo
  {author} {\bibfnamefont {R.}~\bibnamefont {{Spurzem}}}, \bibinfo {author}
  {\bibfnamefont {W.}~\bibnamefont {{Tang}}}, \bibinfo {author} {\bibfnamefont
  {Y.}~\bibnamefont {{Wang}}}, \bibinfo {author} {\bibfnamefont
  {Y.}~\bibnamefont {{Wang}}}, \bibinfo {author} {\bibfnamefont
  {Y.}~\bibnamefont {{Zang}}}, \ and\ \bibinfo {author} {\bibfnamefont
  {Y.}~\bibnamefont {{Lau}}},\ }\href {\doibase 10.1360/SSPMA2016-00438}
  {\bibfield  {journal} {\bibinfo  {journal} {Scientia Sinica Physica,
  Mechanica \& Astronomica}\ }\textbf {\bibinfo {volume} {47}},\ \bibinfo
  {pages} {010404} (\bibinfo {year} {2017})}\BibitemShut {NoStop}%
\bibitem [{\citenamefont {{Luo}}\ \emph {et~al.}(2016)\citenamefont {{Luo}},
  \citenamefont {{Chen}}, \citenamefont {{Duan}}, \citenamefont {{Gong}},
  \citenamefont {{Hu}}, \citenamefont {{Ji}}, \citenamefont {{Liu}},
  \citenamefont {{Mei}}, \citenamefont {{Milyukov}}, \citenamefont {{Sazhin}},
  \citenamefont {{Shao}}, \citenamefont {{Toth}}, \citenamefont {{Tu}},
  \citenamefont {{Wang}}, \citenamefont {{Wang}}, \citenamefont {{Yeh}},
  \citenamefont {{Zhan}}, \citenamefont {{Zhang}}, \citenamefont {{Zharov}},\
  and\ \citenamefont {{Zhou}}}]{tianqin}%
  \BibitemOpen
  \bibfield  {author} {\bibinfo {author} {\bibfnamefont {J.}~\bibnamefont
  {{Luo}}}, \bibinfo {author} {\bibfnamefont {L.-S.}\ \bibnamefont {{Chen}}},
  \bibinfo {author} {\bibfnamefont {H.-Z.}\ \bibnamefont {{Duan}}}, \bibinfo
  {author} {\bibfnamefont {Y.-G.}\ \bibnamefont {{Gong}}}, \bibinfo {author}
  {\bibfnamefont {S.}~\bibnamefont {{Hu}}}, \bibinfo {author} {\bibfnamefont
  {J.}~\bibnamefont {{Ji}}}, \bibinfo {author} {\bibfnamefont {Q.}~\bibnamefont
  {{Liu}}}, \bibinfo {author} {\bibfnamefont {J.}~\bibnamefont {{Mei}}},
  \bibinfo {author} {\bibfnamefont {V.}~\bibnamefont {{Milyukov}}}, \bibinfo
  {author} {\bibfnamefont {M.}~\bibnamefont {{Sazhin}}}, \bibinfo {author}
  {\bibfnamefont {C.-G.}\ \bibnamefont {{Shao}}}, \bibinfo {author}
  {\bibfnamefont {V.~T.}\ \bibnamefont {{Toth}}}, \bibinfo {author}
  {\bibfnamefont {H.-B.}\ \bibnamefont {{Tu}}}, \bibinfo {author}
  {\bibfnamefont {Y.}~\bibnamefont {{Wang}}}, \bibinfo {author} {\bibfnamefont
  {Y.}~\bibnamefont {{Wang}}}, \bibinfo {author} {\bibfnamefont {H.-C.}\
  \bibnamefont {{Yeh}}}, \bibinfo {author} {\bibfnamefont {M.-S.}\ \bibnamefont
  {{Zhan}}}, \bibinfo {author} {\bibfnamefont {Y.}~\bibnamefont {{Zhang}}},
  \bibinfo {author} {\bibfnamefont {V.}~\bibnamefont {{Zharov}}}, \ and\
  \bibinfo {author} {\bibfnamefont {Z.-B.}\ \bibnamefont {{Zhou}}},\ }\href
  {\doibase 10.1088/0264-9381/33/3/035010} {\bibfield  {journal} {\bibinfo
  {journal} {Classical and Quantum Gravity}\ }\textbf {\bibinfo {volume}
  {33}},\ \bibinfo {eid} {035010} (\bibinfo {year} {2016})}\BibitemShut
  {NoStop}%
\bibitem [{\citenamefont {Afzal}\ \emph {et~al.}(2023)\citenamefont {Afzal}
  \emph {et~al.}}]{NANOGrav_1}%
  \BibitemOpen
  \bibfield  {author} {\bibinfo {author} {\bibfnamefont {A.}~\bibnamefont
  {Afzal}} \emph {et~al.},\ }\href {\doibase 10.3847/2041-8213/acdc91}
  {\bibfield  {journal} {\bibinfo  {journal} {The Astrophysical Journal
  Letters}\ }\textbf {\bibinfo {volume} {951}},\ \bibinfo {pages} {L11}
  (\bibinfo {year} {2023})}\BibitemShut {NoStop}%
\bibitem [{\citenamefont {Afzal}\ \emph {et~al.}(2024)\citenamefont {Afzal}
  \emph {et~al.}}]{NANOGrav_2}%
  \BibitemOpen
  \bibfield  {author} {\bibinfo {author} {\bibfnamefont {A.}~\bibnamefont
  {Afzal}} \emph {et~al.},\ }\href {\doibase 10.3847/2041-8213/ad68fc}
  {\bibfield  {journal} {\bibinfo  {journal} {The Astrophysical Journal
  Letters}\ }\textbf {\bibinfo {volume} {971}},\ \bibinfo {pages} {L27}
  (\bibinfo {year} {2024})}\BibitemShut {NoStop}%
\bibitem [{\citenamefont {Caprini}\ \emph {et~al.}(2016)\citenamefont
  {Caprini}, \citenamefont {Hindmarsh}, \citenamefont {Huber}, \citenamefont
  {Konstandin}, \citenamefont {Kozaczuk}, \citenamefont {Nardini},
  \citenamefont {No}, \citenamefont {Petiteau}, \citenamefont {Schwaller},
  \citenamefont {Servant},\ and\ \citenamefont {Weir}}]{SGWB_LISA_1}%
  \BibitemOpen
  \bibfield  {author} {\bibinfo {author} {\bibfnamefont {C.}~\bibnamefont
  {Caprini}}, \bibinfo {author} {\bibfnamefont {M.}~\bibnamefont {Hindmarsh}},
  \bibinfo {author} {\bibfnamefont {S.}~\bibnamefont {Huber}}, \bibinfo
  {author} {\bibfnamefont {T.}~\bibnamefont {Konstandin}}, \bibinfo {author}
  {\bibfnamefont {J.}~\bibnamefont {Kozaczuk}}, \bibinfo {author}
  {\bibfnamefont {G.}~\bibnamefont {Nardini}}, \bibinfo {author} {\bibfnamefont
  {J.~M.}\ \bibnamefont {No}}, \bibinfo {author} {\bibfnamefont
  {A.}~\bibnamefont {Petiteau}}, \bibinfo {author} {\bibfnamefont
  {P.}~\bibnamefont {Schwaller}}, \bibinfo {author} {\bibfnamefont
  {G.}~\bibnamefont {Servant}}, \ and\ \bibinfo {author} {\bibfnamefont
  {D.~J.}\ \bibnamefont {Weir}},\ }\href {\doibase
  10.1088/1475-7516/2016/04/001} {\bibfield  {journal} {\bibinfo  {journal}
  {Journal of Cosmology and Astroparticle Physics}\ }\textbf {\bibinfo {volume}
  {2016}},\ \bibinfo {pages} {001} (\bibinfo {year} {2016})}\BibitemShut
  {NoStop}%
\bibitem [{\citenamefont {Caprini}\ \emph {et~al.}(2020)\citenamefont
  {Caprini}, \citenamefont {Chala}, \citenamefont {Dorsch}, \citenamefont
  {Hindmarsh}, \citenamefont {Huber}, \citenamefont {Konstandin}, \citenamefont
  {Kozaczuk}, \citenamefont {Nardini}, \citenamefont {No}, \citenamefont
  {Rummukainen}, \citenamefont {Schwaller}, \citenamefont {Servant},
  \citenamefont {Tranberg},\ and\ \citenamefont {Weir}}]{SGWB_LISA_2}%
  \BibitemOpen
  \bibfield  {author} {\bibinfo {author} {\bibfnamefont {C.}~\bibnamefont
  {Caprini}}, \bibinfo {author} {\bibfnamefont {M.}~\bibnamefont {Chala}},
  \bibinfo {author} {\bibfnamefont {G.~C.}\ \bibnamefont {Dorsch}}, \bibinfo
  {author} {\bibfnamefont {M.}~\bibnamefont {Hindmarsh}}, \bibinfo {author}
  {\bibfnamefont {S.~J.}\ \bibnamefont {Huber}}, \bibinfo {author}
  {\bibfnamefont {T.}~\bibnamefont {Konstandin}}, \bibinfo {author}
  {\bibfnamefont {J.}~\bibnamefont {Kozaczuk}}, \bibinfo {author}
  {\bibfnamefont {G.}~\bibnamefont {Nardini}}, \bibinfo {author} {\bibfnamefont
  {J.~M.}\ \bibnamefont {No}}, \bibinfo {author} {\bibfnamefont
  {K.}~\bibnamefont {Rummukainen}}, \bibinfo {author} {\bibfnamefont
  {P.}~\bibnamefont {Schwaller}}, \bibinfo {author} {\bibfnamefont
  {G.}~\bibnamefont {Servant}}, \bibinfo {author} {\bibfnamefont
  {A.}~\bibnamefont {Tranberg}}, \ and\ \bibinfo {author} {\bibfnamefont
  {D.~J.}\ \bibnamefont {Weir}},\ }\href {\doibase
  10.1088/1475-7516/2020/03/024} {\bibfield  {journal} {\bibinfo  {journal}
  {Journal of Cosmology and Astroparticle Physics}\ }\textbf {\bibinfo {volume}
  {2020}},\ \bibinfo {pages} {024} (\bibinfo {year} {2020})}\BibitemShut
  {NoStop}%
\bibitem [{\citenamefont {Banerjee}\ and\ \citenamefont
  {Pav\'on}(2001)}]{MG_1}%
  \BibitemOpen
  \bibfield  {author} {\bibinfo {author} {\bibfnamefont {N.}~\bibnamefont
  {Banerjee}}\ and\ \bibinfo {author} {\bibfnamefont {D.}~\bibnamefont
  {Pav\'on}},\ }\href {\doibase 10.1103/PhysRevD.63.043504} {\bibfield
  {journal} {\bibinfo  {journal} {Phys. Rev. D}\ }\textbf {\bibinfo {volume}
  {63}},\ \bibinfo {pages} {043504} (\bibinfo {year} {2001})}\BibitemShut
  {NoStop}%
\bibitem [{\citenamefont {Sen}\ and\ \citenamefont {Sen}(2001)}]{MG_2}%
  \BibitemOpen
  \bibfield  {author} {\bibinfo {author} {\bibfnamefont {S.}~\bibnamefont
  {Sen}}\ and\ \bibinfo {author} {\bibfnamefont {A.~A.}\ \bibnamefont {Sen}},\
  }\href {\doibase 10.1103/PhysRevD.63.124006} {\bibfield  {journal} {\bibinfo
  {journal} {Phys. Rev. D}\ }\textbf {\bibinfo {volume} {63}},\ \bibinfo
  {pages} {124006} (\bibinfo {year} {2001})}\BibitemShut {NoStop}%
\bibitem [{\citenamefont {Qiang}\ \emph {et~al.}(2005)\citenamefont {Qiang},
  \citenamefont {Ma}, \citenamefont {Han},\ and\ \citenamefont {Yu}}]{MG_3}%
  \BibitemOpen
  \bibfield  {author} {\bibinfo {author} {\bibfnamefont {L.-e.}\ \bibnamefont
  {Qiang}}, \bibinfo {author} {\bibfnamefont {Y.}~\bibnamefont {Ma}}, \bibinfo
  {author} {\bibfnamefont {M.}~\bibnamefont {Han}}, \ and\ \bibinfo {author}
  {\bibfnamefont {D.}~\bibnamefont {Yu}},\ }\href {\doibase
  10.1103/PhysRevD.71.061501} {\bibfield  {journal} {\bibinfo  {journal} {Phys.
  Rev. D}\ }\textbf {\bibinfo {volume} {71}},\ \bibinfo {pages} {061501}
  (\bibinfo {year} {2005})}\BibitemShut {NoStop}%
\bibitem [{\citenamefont {Henry~Tye}\ and\ \citenamefont
  {Wasserman}(2001)}]{HD}%
  \BibitemOpen
  \bibfield  {author} {\bibinfo {author} {\bibfnamefont {S.-H.}\ \bibnamefont
  {Henry~Tye}}\ and\ \bibinfo {author} {\bibfnamefont {I.}~\bibnamefont
  {Wasserman}},\ }\href {\doibase 10.1103/PhysRevLett.86.1682} {\bibfield
  {journal} {\bibinfo  {journal} {Phys. Rev. Lett.}\ }\textbf {\bibinfo
  {volume} {86}},\ \bibinfo {pages} {1682} (\bibinfo {year}
  {2001})}\BibitemShut {NoStop}%
\bibitem [{\citenamefont {Peebles}\ and\ \citenamefont {Ratra}(2003)}]{CC_1}%
  \BibitemOpen
  \bibfield  {author} {\bibinfo {author} {\bibfnamefont {P.~J.~E.}\
  \bibnamefont {Peebles}}\ and\ \bibinfo {author} {\bibfnamefont
  {B.}~\bibnamefont {Ratra}},\ }\href {\doibase 10.1103/RevModPhys.75.559}
  {\bibfield  {journal} {\bibinfo  {journal} {Rev. Mod. Phys.}\ }\textbf
  {\bibinfo {volume} {75}},\ \bibinfo {pages} {559} (\bibinfo {year}
  {2003})}\BibitemShut {NoStop}%
\bibitem [{\citenamefont {Weinberg}(1989)}]{CC_2}%
  \BibitemOpen
  \bibfield  {author} {\bibinfo {author} {\bibfnamefont {S.}~\bibnamefont
  {Weinberg}},\ }\href {\doibase 10.1103/RevModPhys.61.1} {\bibfield  {journal}
  {\bibinfo  {journal} {Rev. Mod. Phys.}\ }\textbf {\bibinfo {volume} {61}},\
  \bibinfo {pages} {1} (\bibinfo {year} {1989})}\BibitemShut {NoStop}%
\bibitem [{\citenamefont {Abbott}\ \emph {et~al.}(2017)\citenamefont {Abbott}
  \emph {et~al.}}]{Cosmology_1}%
  \BibitemOpen
  \bibfield  {author} {\bibinfo {author} {\bibfnamefont {B.~P.}\ \bibnamefont
  {Abbott}} \emph {et~al.},\ }\href {\doibase 10.1038/nature24471} {\bibfield
  {journal} {\bibinfo  {journal} {Nature}\ }\textbf {\bibinfo {volume} {551}},\
  \bibinfo {pages} {85} (\bibinfo {year} {2017})}\BibitemShut {NoStop}%
\bibitem [{\citenamefont {Holz}\ and\ \citenamefont
  {Linder}(2005)}]{Cosmology_2}%
  \BibitemOpen
  \bibfield  {author} {\bibinfo {author} {\bibfnamefont {D.~E.}\ \bibnamefont
  {Holz}}\ and\ \bibinfo {author} {\bibfnamefont {E.~V.}\ \bibnamefont
  {Linder}},\ }\href {\doibase 10.1086/432085} {\bibfield  {journal} {\bibinfo
  {journal} {The Astrophysical Journal}\ }\textbf {\bibinfo {volume} {631}},\
  \bibinfo {pages} {678} (\bibinfo {year} {2005})}\BibitemShut {NoStop}%
\bibitem [{\citenamefont {Dalal}\ \emph {et~al.}(2006)\citenamefont {Dalal},
  \citenamefont {Holz}, \citenamefont {Hughes},\ and\ \citenamefont
  {Jain}}]{Cosmology_3}%
  \BibitemOpen
  \bibfield  {author} {\bibinfo {author} {\bibfnamefont {N.}~\bibnamefont
  {Dalal}}, \bibinfo {author} {\bibfnamefont {D.~E.}\ \bibnamefont {Holz}},
  \bibinfo {author} {\bibfnamefont {S.~A.}\ \bibnamefont {Hughes}}, \ and\
  \bibinfo {author} {\bibfnamefont {B.}~\bibnamefont {Jain}},\ }\href {\doibase
  10.1103/PhysRevD.74.063006} {\bibfield  {journal} {\bibinfo  {journal} {Phys.
  Rev. D}\ }\textbf {\bibinfo {volume} {74}},\ \bibinfo {pages} {063006}
  (\bibinfo {year} {2006})}\BibitemShut {NoStop}%
\bibitem [{\citenamefont {Valentino}\ \emph {et~al.}(2021)\citenamefont
  {Valentino}, \citenamefont {Mena}, \citenamefont {Pan}, \citenamefont
  {Visinelli}, \citenamefont {Yang}, \citenamefont {Melchiorri}, \citenamefont
  {Mota}, \citenamefont {Riess},\ and\ \citenamefont {Silk}}]{Cosmology_4}%
  \BibitemOpen
  \bibfield  {author} {\bibinfo {author} {\bibfnamefont {E.~D.}\ \bibnamefont
  {Valentino}}, \bibinfo {author} {\bibfnamefont {O.}~\bibnamefont {Mena}},
  \bibinfo {author} {\bibfnamefont {S.}~\bibnamefont {Pan}}, \bibinfo {author}
  {\bibfnamefont {L.}~\bibnamefont {Visinelli}}, \bibinfo {author}
  {\bibfnamefont {W.}~\bibnamefont {Yang}}, \bibinfo {author} {\bibfnamefont
  {A.}~\bibnamefont {Melchiorri}}, \bibinfo {author} {\bibfnamefont {D.~F.}\
  \bibnamefont {Mota}}, \bibinfo {author} {\bibfnamefont {A.~G.}\ \bibnamefont
  {Riess}}, \ and\ \bibinfo {author} {\bibfnamefont {J.}~\bibnamefont {Silk}},\
  }\href {\doibase 10.1088/1361-6382/ac086d} {\bibfield  {journal} {\bibinfo
  {journal} {Classical and Quantum Gravity}\ }\textbf {\bibinfo {volume}
  {38}},\ \bibinfo {pages} {153001} (\bibinfo {year} {2021})}\BibitemShut
  {NoStop}%
\bibitem [{\citenamefont {Sasaki}\ \emph {et~al.}(2016)\citenamefont {Sasaki},
  \citenamefont {Suyama}, \citenamefont {Tanaka},\ and\ \citenamefont
  {Yokoyama}}]{Misao_16}%
  \BibitemOpen
  \bibfield  {author} {\bibinfo {author} {\bibfnamefont {M.}~\bibnamefont
  {Sasaki}}, \bibinfo {author} {\bibfnamefont {T.}~\bibnamefont {Suyama}},
  \bibinfo {author} {\bibfnamefont {T.}~\bibnamefont {Tanaka}}, \ and\ \bibinfo
  {author} {\bibfnamefont {S.}~\bibnamefont {Yokoyama}},\ }\href {\doibase
  10.1103/PhysRevLett.117.061101} {\bibfield  {journal} {\bibinfo  {journal}
  {Phys. Rev. Lett.}\ }\textbf {\bibinfo {volume} {117}},\ \bibinfo {pages}
  {061101} (\bibinfo {year} {2016})}\BibitemShut {NoStop}%
\bibitem [{\citenamefont {N\"af}\ \emph {et~al.}(2009)\citenamefont {N\"af},
  \citenamefont {Jetzer},\ and\ \citenamefont {Sereno}}]{Lambda_09}%
  \BibitemOpen
  \bibfield  {author} {\bibinfo {author} {\bibfnamefont {J.}~\bibnamefont
  {N\"af}}, \bibinfo {author} {\bibfnamefont {P.}~\bibnamefont {Jetzer}}, \
  and\ \bibinfo {author} {\bibfnamefont {M.}~\bibnamefont {Sereno}},\ }\href
  {\doibase 10.1103/PhysRevD.79.024014} {\bibfield  {journal} {\bibinfo
  {journal} {Phys. Rev. D}\ }\textbf {\bibinfo {volume} {79}},\ \bibinfo
  {pages} {024014} (\bibinfo {year} {2009})}\BibitemShut {NoStop}%
\bibitem [{\citenamefont {Khan}\ \emph {et~al.}(2016)\citenamefont {Khan},
  \citenamefont {Husa}, \citenamefont {Hannam}, \citenamefont {Ohme},
  \citenamefont {P\"urrer}, \citenamefont {Forteza},\ and\ \citenamefont
  {Boh\'e}}]{PhD}%
  \BibitemOpen
  \bibfield  {author} {\bibinfo {author} {\bibfnamefont {S.}~\bibnamefont
  {Khan}}, \bibinfo {author} {\bibfnamefont {S.}~\bibnamefont {Husa}}, \bibinfo
  {author} {\bibfnamefont {M.}~\bibnamefont {Hannam}}, \bibinfo {author}
  {\bibfnamefont {F.}~\bibnamefont {Ohme}}, \bibinfo {author} {\bibfnamefont
  {M.}~\bibnamefont {P\"urrer}}, \bibinfo {author} {\bibfnamefont {X.~J.}\
  \bibnamefont {Forteza}}, \ and\ \bibinfo {author} {\bibfnamefont
  {A.}~\bibnamefont {Boh\'e}},\ }\href {\doibase 10.1103/PhysRevD.93.044007}
  {\bibfield  {journal} {\bibinfo  {journal} {Phys. Rev. D}\ }\textbf {\bibinfo
  {volume} {93}},\ \bibinfo {pages} {044007} (\bibinfo {year}
  {2016})}\BibitemShut {NoStop}%
\bibitem [{\citenamefont {Maggiore}(2007)}]{Maggiore2007}%
  \BibitemOpen
  \bibfield  {author} {\bibinfo {author} {\bibfnamefont {M.}~\bibnamefont
  {Maggiore}},\ }\href {\doibase 10.1093/acprof:oso/9780198570745.001.0001}
  {\emph {\bibinfo {title} {{Gravitational Waves: Volume 1: Theory and
  Experiments}}}}\ (\bibinfo  {publisher} {Oxford University Press},\ \bibinfo
  {year} {2007})\BibitemShut {NoStop}%
\bibitem [{\citenamefont {Kaiser}\ and\ \citenamefont
  {McWilliams}(2021)}]{gwent}%
  \BibitemOpen
  \bibfield  {author} {\bibinfo {author} {\bibfnamefont {A.~R.}\ \bibnamefont
  {Kaiser}}\ and\ \bibinfo {author} {\bibfnamefont {S.~T.}\ \bibnamefont
  {McWilliams}},\ }\href {\doibase 10.1088/1361-6382/abd4f6} {\bibfield
  {journal} {\bibinfo  {journal} {Classical and Quantum Gravity}\ }\textbf
  {\bibinfo {volume} {38}},\ \bibinfo {pages} {055009} (\bibinfo {year}
  {2021})},\ \Eprint {http://arxiv.org/abs/2010.02135} {2010.02135}
  \BibitemShut {NoStop}%
\bibitem [{\citenamefont {Cutler}\ and\ \citenamefont {Flanagan}(1994)}]{FIM}%
  \BibitemOpen
  \bibfield  {author} {\bibinfo {author} {\bibfnamefont {C.}~\bibnamefont
  {Cutler}}\ and\ \bibinfo {author} {\bibfnamefont {E.~E.}\ \bibnamefont
  {Flanagan}},\ }\href {\doibase 10.1103/PhysRevD.49.2658} {\bibfield
  {journal} {\bibinfo  {journal} {Phys. Rev. D}\ }\textbf {\bibinfo {volume}
  {49}},\ \bibinfo {pages} {2658} (\bibinfo {year} {1994})}\BibitemShut
  {NoStop}%
\end{thebibliography}

\end{document}